# STOCK BROAD-INDEX TREND PATTERNS LEARNING VIA DOMAIN KNOWLEDGE INFORMED GENERATIVE NETWORK

Jingyi Gu, Fadi P. Deek, and Guiling Wang

New Jersey Institute of Technology, Newark, NJ, USA
jg95@njit.edu, Deek@njit.edu, and gwang@njit.edu

## ABSTRACT

*Predicting the Stock movement attracts much attention from both industry and academia. Despite such significant efforts, the results remain unsatisfactory due to the inherently complicated nature of the stock market driven by factors including supply and demand, the state of the economy, the political climate, and even irrational human behavior. Recently, Generative Adversarial Networks (GAN) have been extended for time series data; however, robust methods are primarily for synthetic series generation, which fall short for appropriate stock prediction. This is because existing GANs for stock applications suffer from mode collapse and only consider one-step prediction, thus underutilizing the potential of GAN. Furthermore, merging news and market volatility are neglected in current GANs. To address these issues, we exploit expert domain knowledge in finance and, for the first time, attempt to formulate stock movement prediction into a Wasserstein GAN framework for multi-step prediction. We propose IndexGAN, which includes deliberate designs for the inherent characteristics of the stock market, leverages news context learning to thoroughly investigate textual information and develop an attentive seq2seq learning network that captures the temporal dependency among stock prices, news, and market sentiment. We also utilize the critic to approximate the Wasserstein distance between actual and predicted sequences and develop a rolling strategy for deployment that mitigates noise from the financial market. Extensive experiments are conducted on real-world broad-based indices, demonstrating the superior performance of our architecture over other state-of-the-art baselines, also validating all its contributing components.*

## 1. INTRODUCTION

The stock market is an essential component of a broad and intricate financial system, and stock prices reflect the dynamics of economic and financial activities. Predicting future movements of either individual stocks or overall market indices is important to investors and other market players [1], which requires significant efforts but lacks satisfactory results. Conventional approaches vary from fundamental and technical analysis to linear statistical models, such as Momentum Strategies and Autoregressive Integrated Moving Average (ARIMA), which capture simple short-term patterns from historical prices. With the tremendous power and success in exploring the nonlinear relationship and dealing with big data, machine learning and neural networks are increasingly utilized in stock movement prediction and have shown better results in prediction accuracy over traditional methods [2].

However, stock prices, especially broad-based indices (such as Dow30 and S&P 500), carry inherent stochasticity caused by crowd emotional factors (such as fear and greed), economic factors (such as fiscal and financial policy and economic health), and many other unknown factors [3]. Thus, existing models suffer from a lack of robustness, and the prediction accuracy is still far from being satisfactory. Additionally, unlike many other tasks, the prediction accuracy of broad-index return cannot keep increasing, in theory, because market arbitrage opportunities will degrade accuracy over time. Naturally, an effective trading strategy that is likely to generate profit will be adopted by an increasing number of investors; eventually, such a strategy may no longer perform as well considering that the market is composed of its collective participants. Thus, in practice, if a certain method is capable of achieving over 60% accuracy using a dataset of a lengthy period, encompassing a financial crisis, it is then considered powerful enough. Therefore, we aim to provide a framework that performs with accuracy, over time, and in different market conditions.

Recently, a powerful machine learning framework named Generative Adversarial Networks (GAN) [4] achieved noteworthy success in various domains. Distinct from single neural networks, GAN trains two models contesting with each other simultaneously, a generator capturing data distribution and a discriminator estimating the probability of a sample from training data instead of a generator. Having achieved promising performance in image generation, researchers have extended GANs to the field of time series for applications in synthetic time series data generation. While aiming to simulate characteristics in sequences, these efforts have rarely been practical enough to **predict** future patterns. Although some advanced methods have been proposed to forecast stock pricing by building an architecture of GAN conditioned on historical prices [5, 6], they have two main deficiencies: 1) Their implemented vanilla GAN suffers from mode collapse problem [7] and creates samples with low diversity. Consequently, most predictions tend to be biased towards an up-market. 2) Existing models only predict one future step, which is then concatenated with historical prices to generate fake sequences. As a result, the only difference between the actual and generated sequence is the last element, and thus the discriminator can easily ignore it. Such methods largely underutilize GAN's potential, which can create sequences with periodical patterns of the stock market. We aim to address these issues by fully realizing the potential of GAN to solve the stock index prediction problem.

In addition to historical prices, other driving factors such as emerging news and market volatility should also be exploited in market prediction. An increasing amount of research results has shown that the stock market is highly affected by significant events [8]. With the popularity of social media, frequent financial reporting, targeted press releases, and news articles significantly influence investors' decisions and drive stock price movement. Natural Language Processing (NLP) techniques have been utilized to analyze textual information to forecast stock pricing. In addition to news, the volatility index, derived from options prices, contains important indicators on market sentiment [9, 10, 11]. There are some common similarity between options and insurance. For example, put options can be purchased as insurance, or a hedge, against a sudden stock price drop. A price increase in put options with the same underlying stock price indicates that the market expects a high probability of price drop, just as an increased flood insurance price imply a higher probability of flood though people may still choose to stay in their homes (i.e., to keep the stock position unsold). The volatility index, which is derived from the prices of index options with near-term expiration dates, is an important leading indicator of the stock market but is often neglected in the GAN models for predicting the stock market. We incorporate both news and volatility index in our proposed prediction model.

Because existing efforts predominantly rely on historical prices and neglect the inherent characteristics of the stock market, there is a need to better address the issues and yield higher prediction accuracy. Thus, we propose the following strategies which employ domain knowledge in financial systems: 1) formulating stock movement to a multiple-step prediction problem; 2) exploiting Wasserstein GAN to avoid mode collapse that tends to predict an up-market, and improve stability by prior noise; 3) leveraging volatility index to capture the market sentiment of seasoned investors; 4) designing news context learning to fully explore textual information from public news and events; and 5) adopting a rolling strategy for deployment to mitigate noise from uncertainty in the financial market, and generate the prediction with horizon-wise information.

Specifically, our proposed *IndexGAN* contains both a generator and a critic to predict the stock movement. Inside the generator, a news context learning module extracts latent word representations from GloVe embedding matrix of news for two purposes. Firstly, it fully explores context information from the news. Secondly, it strategically shrinks the size of the embedding matrix to a comparable extent with price features, considering the size of GloVe embedding matrix is several orders of magnitude larger than market-based features and prior noise, and simply concatenating them together will let the following seq2seq learning entirely concentrate on news and mask the impact of other features. An attentive seq2seq learning module captures temporal dependence from historical prices, volatility index, various technical indicators, prior noise, and latent word representations. The generator with both modules predicts the future sequence. The critic (discriminator in GAN) is designed to approximate the Wasserstein distance between real and generated sequences, which is incorporated into an adversarial loss for optimization. During the training process, besides adversarial loss, two additional losses are added to improve the similarity between real and fake distribution efficiently and avoid biased movement.

Accurate predictions of an index return enable hedging against black swan events and thus can significantly reduce risk and help maintain market liquidity when facing extreme circumstances, thereby contributing to financial market stabilization. *IndexGAN* is implemented on broad indices. To understand the complexity and dynamics of the real-world market index, we select a period of high volatility that covers the 2009 financial crisis and the subsequent recovery for training and testing of *IndexGAN*. Experiment results show that the model boosts the accuracy from around 58% to over 60% for both indices (Dow Jones Industrial Average (DJIA) and S&P 500) compared to the previous state-of-the-art model. This improvement demonstrates the effectiveness of the proposed method. The major contributions of this work are summarized as follows:

1. To the best of our knowledge, ours is a first attempt at formulating stock movement predictions into a Wasserstein generative adversarial networks framework for multi-step prediction. The new formulation enhances the robustness by prior noise and provides reasonable guidance on market timing to investors through the use of multi-step prediction.

2. *IndexGAN* consists of particularly crafted designs that consider the inherent characteristics of the stock market. It includes a news context learning module, features of volatility index and multiple empirically valid technical indicators, and a rolling deployment strategy to mitigate uncertainty and generate prediction with horizon-wise information.

3. We conduct extensive experiments on two stock indices that attract the most global attention. *IndexGAN* achieves consistent and significant improvements over multiple popular benchmarks.

## 2. RELATED WORKS

### 2.1. Stock Market Prediction

Stock movement prediction has drawn much industry attention for its potential to maximize profit, as has the intricate nature of this problem to academia. Traditional methods fall into two categories: fundamental analysis and technical analysis [12]. Fundamental analysis attempts to evaluate the fair and long-term stock price of a firm by examining its fundamental status and comparative strength within its industry as well as the overall economy such as revenues, earnings, and account payables [13, 14]. Technical analysis, also known as chart analysis, examines price trends and identifies price patterns [15]. There are numerous technical indicators, such as moving average and Bollinger Bands, that guide traders on buying and selling trends. Linear models, such as ARIMA [16] and Generalized Autoregressive Conditional Heteroskedasticity (GARCH) [17], have also been utilized in stock prediction. With the development of deep learning, many deep neural networks have been employed in the finance area, especially for stock prediction [18]. Qin et al. [19] propose a dual-stage attention-based Recurrent Neural Network (RNN) model to predict stock trends. Feng et al. [20] further consider employing the concept of adversarial training on Long Short-Term Memory (LSTM) [21] to predict stock market pricing. Ding et al. [22] enhance the transformer with trading gap splitter and gaussian prior for stock price prediction.

As the outreach of digital and social media grow enabling real-time streaming of financial news, their impact on the stock market is explored. Existing work for textual stock analysis can be classified into sentiment analysis and context-aware approaches. Sentiment analysis is leveraged to extract emotions from text and generate sentiment scores as features [23], although this often ignores useful textual information. Zhang, Fuehres, and Gloor [24] use mood words as emotional tags to classify the sentiment of tweets. De Fortuny et al. [25] select sentiment words from the news by TF-IDF and calculate sentiment polarity and subjectivity as sentiment features for Support Vector Machine (SVM) in making predictions. Zhao et al. [26] employ LDA to filter microblogs and create a financial sentiment lexicon to extract public emotions. On the other hand, context-aware approaches create embeddings from news and feed them into complicated structures to explore context as much as possible [27, 28, 29], which often suffer from the overfitting problem. Lee and Soo [30] process financial news with word2vec to create input for recurrent Convolutional Neural Network (CNN). Minh et al. [31] implement Stock2Vec and two-stream Gated Recurrent Unit (GRU) models to create embeddings from financial news for stock price classification. Li et al. [32] propose using Graph Convolutional Network to represent connection given news embeddings and predict overnight movement. Both methods are widely used for leveraging news in stock market prediction.

Although existing efforts have been successful in achieving their stated goals, market risk is still largely not adequately addressed. Hence, we aim to incorporate market volatility into stock prediction.

## 2.2. Generative Adversarial Nets on Time Series

More recently, there are emerging efforts in using GAN on time series data in the domain of video generation [33] in medicine [34], text generation [35], biology [36], finance [37], and music [38]. These GANs aim to generate a synthetic sequence that retains the characteristics of the original sequence. For instance, Yu et al. [39] introduce SeqGAN, which extends GANs with reinforcement learning to generate sequences of discrete tokens. Yoon et al. [40] propose TimeGAN for realistic time-series generation, combining an unsupervised paradigm with supervised autoregressive models. These methods are used for tasks of enriching limited real-world datasets [41], data imputation [42], and machine translation [43]. However, stock market predictions are attempts to discover future patterns instead of simulating original time series and thus cannot employ most existing efforts on time series GANs.

Some recent approaches implement GANs to predict stock trends. Zhang et al. [44] employ LSTM as a generator and MLP as a discriminator to forecast stock pricing. Faraz and Khaloozadeh [6] enhance GAN with wavelet transformation and z-score method in the data preprocessing phase. Sonkiya et al. [45] generate binary sentiment score by FinBERT and employ GAN for stock price predictions. Such methods adapt Vanilla GAN which often suffers from mode collapse, and neglect possible driving factors such as complex public emotions, financial policies, and events, which limits the ability to achieve superior performance. In this work, we use GAN with Wasserstein distance to design a robust architecture enabling us to capture the characteristics of the stock market.

## 3. PROBLEM FORMULATION

We aim to develop a model that accurately predicts whether a stock index is up or down the next day. If the close price of the next day is higher than that of the current day, we have an up day and vice versa. We select two different types of features: market-based features read derived from the stock market data and news-based features derived from raw news headlines. Market-based features are listed in Table 1. Specifically, we retrieve the open, high, low, and close prices of the stock index on trading days. We also deliberately select nine empirically useful technical indicators, calculate their values based on the price data, and employ them as features. In addition, we employ the market volatility data, i.e., VIX, if we predict the S&P 500 index and VXD if we predict the Dow Jones Index.

Let $\mathcal{S} = \{s_t\}_{t=1}^n \in \mathbb{R}^{n \times p}$ be a matrix which represents the stock dataset with $n$ time points, where $s_t$ is a vector with $p$ market-based features. Let $\mathcal{H} = \{h_t\}_{t=1}^n \in \mathbb{R}^{n \times k}$ be a matrix which denotes textual dataset, we assume that each vector $h_t$ contains a number of $k$ raw news headlines at the time step $t$. Let $z$ denote the prior noise which is assumed to be drawn from Gaussian distribution. We concatenate stock features and headlines together. Then we denote our training data as $\mathcal{D} = \{(s_t, h_t, z)\}_{t=1}^n$. To investigate the temporal dependency of sequential data, we set the length of historical sequence as $w$ and the length of future sequence as $q$. This means that we use the past $w$ days' data to predict the next $q$ days sequence in the inference phase. Suppose the current time is day $T$, given the trained model, we use $X = \{(s_t, h_t, z)\}_{t=T-w+1}^T$ as input and generate $\hat{c} = \{\hat{c}_t\}_{t=T+1}^{T+q}$, where $\hat{c}_t$ is the predicted return based on close price. From the generated sequence, we can determine whether day $T + 1$ is up or down.

Table 1. Market-based features. $s_t$ denotes corresponding features before processing. BOLU and BOLD are upper and lower Bollinger Bands.

| Type | Features | Calculation |
|---|---|---|
| Price | Open, Close, High, Low | $\frac{s_t}{s_{t-1}} - 1$ |
| Volatility | VIX, VXD | |
| Technical Indicator | RSI | -1 if $s_t \leq 20$, 1 if $s_t \geq 80$, 0 otherwise |
| | MACD DIFF | min-max scaling |
| | Upper Bollinger band indicator | if $close_t > BOLU$, 0 otherwise |
| | Lower Bollinger band indicator | if $close_t < BOLD$, 0 otherwise |
| | EMA: 5-day | $\frac{s_t}{close_t} - 1$ |
| | SMA: 13-day, 21-day, 50-day, 200-day | |

## 4. METHODOLOGIES

We elect to use GAN, specifically *IndexGAN*, for stock market prediction. The architecture of the model is shown in Figure 1. In general, GAN consists of two models: a generator that learns distribution over real data, and a discriminator that distinguishes between real data and fake output from the generator. For the purpose of controlling on modes of data being generated, conditional GAN [46] was proposed by conditioning the model to additional information. In our model, stock and news features are used as additional information. The goal is to use training data $\mathcal{D}$ to learn a feature map (generator) $\mathcal{G}$ which best generates sequences of future stock prices in the GAN framework. WGAN [47], which uses Wasserstein distance to measure the difference between the real data distribution and generated data distribution, is chosen for its stability.

### 4.1. Generator

Generator $\mathcal{G}$ with parameter set $\Theta_g$ takes input $X$ and aims to build a fake sequence $\hat{c}$ whose distribution is as close as possible to the real sequence $c = \{c_t\}_{t=T+1}^{T+q}$. Our generator has two parts: news context learning and attentive seq2seq learning.

#### 4.1.1. News Context Learning

To convert the text from news into word vectors, word2vec is applied to encode the textual input as a continuous distributed representation. GloVe, one of the most popular word representation techniques in word2vec, is chosen to be employed in our model. Compared with general word2vec models, GloVe not only considers words and related local context information but also incorporates aggregated global word co-occurrence statistics from the corpus to create an embedding matrix. Given a paragraph $h_t$ at time step $t$ consisting of $k$ news headlines, each word in the headlines is mapped to a vector in $m$ dimensions. The process is formulated as:

$$h'_t = f_{GloVe}(h_t) \qquad (1)$$

where $f_{GloVe}$ is the mapping function from GloVe to word vectors, $h'_t \in \mathbb{R}^{k \times l \times m}$ represents the word embedding matrix, and $l$ is the maximum length of headlines in the textual dataset $\mathcal{H}$. Since the length of words in different headlines may be different, we add zero paddings to the end of word vectors from GloVe if the length is shorter than $l$.

Then a block of nonlinear layers for the embedding matrix is designed. The first component in a block is a fully connected layer. Next, batch normalization is applied to accelerate the training process by normalizing the features in the embedding matrix. In addition, since batch normalization is shown effective in preventing the mode collapse problem, we apply it to all blocks except the last one. The third layer is LeakyReLU activation which avoids saturation in the learning process and dying neurons, except that the last block uses tanh activation function for the output layer. We feed the embedding matrix $h'_t$ into a number of $m$ successive blocks formulated as follows:

$$h_t^1 = \phi(f_{BN}(W_{b1} h'_t + b_{b1})) \qquad (2)$$

$$h_t^2 = \phi(f_{BN}(W_{b2} h_t^1 + b_{b2})) \qquad (3)$$

$$\dots$$

$$\widetilde{h_t} = tanh(W_{bm} h_t^{m-1} + b_{bm}) \qquad (4)$$

where $h_t^i, i = 1, \dots, m$ is the output from $i^{th}$ block and input for $i + 1^{th}$ block, $\widetilde{h_t} \in \mathbb{R}^g$ is the latent word representation with $g$ dimensions, $\phi$ is the LeakyReLU activation function, $W_{bi}$ and $b_{bi}$ are parameters in a fully connected layer.

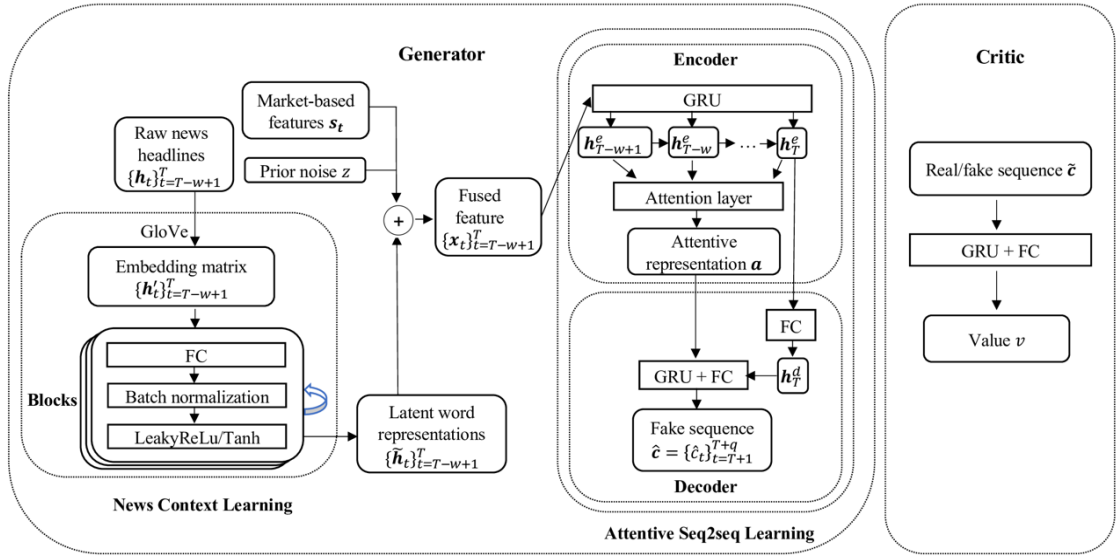

Figure 1 The Proposed *IndexGAN* Structure

### 4.1.2. Attentive Seq2seq Learning

Stock data shows seasonality. For example, a high open price on Monday morning frequently ends up with an up day on Friday. To adaptively capture the temporal relationship and generate sequences, we employ an encoder-decoder with the attentive mechanism as the sequence-to-sequence learning. To improve the robustness of prediction, we add prior noise as an input feature to simulate the stochasticity nature of stock prices. In addition, among market-based features, Exponential Moving Average (EMA) and Simple Moving Average (SMA) over different periods of time (5, 13, 21, 50, and 200 days) are used to capture temporal information across multi-scale time spans.

The encoder extracts the attentive variable $a$. Its input is the sequence of a concatenation of features $x_t$ containing prior noise $z$, market-based feature $s_t$, and latent word representation $\widetilde{h}_t$:

$$x_t = [z, s_t, \widetilde{h}_t] \tag{5}$$

We adopt the simple yet effective GRU to obtain dynamic changes from historical sequence $\{x_t\}_{t=T-w+1}^{T}$:

$$h_t^e = f_e(x_t, h_{t-1}^e) \tag{6}$$

where $f_e$ represents a GRU layer, and $h_{t-1}^e$ denotes the hidden state at each time step.

To adaptively learn the different contributions of input features in the different historical time steps, we implement an attention layer to aggregate information of hidden state from GRU:

$$a_t = v^\top tanh(W_a h_t^e + b_a), \widetilde{a}_t = \frac{exp(a_t)}{\sum_{t=T-w+1}^{T} exp(a_t)} \tag{7}$$

$$a = \sum_{t=T-w+1}^{T} \widetilde{a}_t h_t^e \tag{8}$$

where $\widetilde{a}_t$ denotes contribution score for hidden state $h_t^e$ at time $t$, $v$, $W_a$ and $b_a$ are parameters, and $a$ is the attentive representation which encloses periodical information of the entire sequence.

The decoder aims to extract patterns from the attentive representation $a$ and generate the fake sequence $\hat{c} \in \mathbb{R}^q$ in the future $q$ time steps. We also adopt GRU as the decoder. The hidden state in the decoder $h_T^d$ is transformed from the last hidden state in encoder $h_T^e$ by a fully connected layer. Then the attentive representation $a$ is presented as input of the GRU. The process is computed as follows:

$$h_T^d = tanh(W_d h_T^e + b_d) \tag{9}$$

$$\hat{c} = f_{fc}\left(f_d(a, h_T^d)\right) \tag{10}$$

where $f_d$ denotes the GRU layer in decoder, $W_d$ and $b_d$ are weight and bias parameters, and $f_{fc}$ represents the prediction layer to generate the fake sequence $\hat{c}$.

For training purposes, the next step is to feed both fake and real sequences into critic for further optimization.

### 4.2. Critic

The critic is designed to approximate Wasserstein distance $W(P_r, P_g)$ between real data distribution $P_r$ and generated data distribution $P_g$:

$$W(P_r, P_g) = \inf_{\gamma \in \Pi(P_r, P_g)} E_{(x,y) \sim \gamma}[||x - y||] \tag{11}$$

where $\Pi(P_r, P_g)$ denotes the set of all joint distribution $\gamma(x, y)$ whose marginals are $P_r$ and $P_g$ respectively. The joint distribution $\gamma(x, y)$ represents the mass caused by transporting $x$ to $y$ when transforming $P_r$ to $P_g$. Therefore, Wasserstein distance indicates the cost of such optimal transportation problem in intuition [47,48].

Let $\tilde{c}$ denote the input sequence for either real or fake data. We use GRU to temporally capture the dependency and a fully connected layer to pass the last hidden state, which is presented as:

$$v = f_c(\tilde{c}; \Theta_c) = f_v(f_{GRU}(\tilde{c}))  \quad (12)$$

where $f_c$ denotes the critic consisting of the GRU layer $f_{GRU}$ and the fully connected layer $f_c$, $\Theta_c$ represents the set of parameters in $f_c$, and the output $v$ is the approximated distance value.

## 4.3. Optimization

As the infimum in $W(P_r, P_g)$ is highly intractable, Kantorovich-Rubinstein duality is used to reconstruct the WGAN value function in critic and solve the optimization problem in practice, as follows:

$$\min_{\mathcal{G}} \max_{f_c \in \mathcal{F}} \mathbb{E}_{c \sim P_r}[f_c(c)] - \mathbb{E}_{\hat{c} \sim P_g}[f_c(\hat{c})] \quad (13)$$

where $P_g$ is implicitly defined by $\hat{c}$, $\mathcal{F}$ is the set of 1-Lipschitz functions. The generator is responsible for minimizing the value function in critic which is equivalent to minimize $W(P_r, P_g)$, so that the generated sequence is similar to the real sequence. The critic aims to find the optimal value function to better approximate the Wasserstein Distance.

Relying solely on adversarial feedback is insufficient for the generator to capture the conditional distribution of data. To improve the similarity between two distributions more efficiently, we introduce two additional losses on the generator to discipline the learning. Firstly, the supervised loss $\mathcal{L}_S$ is to directly describe the $L1$ distance between $c$ and $\hat{c}$:

$$\mathcal{L}_S = \mathbb{E}_{\hat{c} \sim P_g, c \sim P_r}[\| c - \hat{c} \|] \quad (14)$$

Secondly, from the perspective of stock movement, the long-term stock market shows an increasing trend, hence the generator tends to construct sequences of higher return even with the detection of short-term fluctuations, which leads to the abnormally high false positive ratio in the classification performance. To address this problem, we design a weighted loss $\mathcal{L}_W$ which represents the error rate with more focus on False Positive (FP) samples than False Negative (FN) samples:

$$\mathcal{L}_W = \frac{1}{q}\left(\lambda_1 \sum_{t=1}^{q} I(y_t < 0, \hat{y}_t > 0) + \lambda_2 \sum_{i=1}^{q} I(y_t > 0, \hat{y}_t < 0)\right) \quad (15)$$

$$\lambda_1 > \lambda_2, \lambda_1 + \lambda_2 = 1, \lambda_1 \in (0,1), \lambda_2 \in (0,1)$$

where $I(\cdot)$ is the indicator function, the movement $y_t = sign(c_t)$ is derived from the return, similar with the predicted movement $\hat{y}_t = sign(\hat{c}_t)$, $\lambda_1$ and $\lambda_2$ are hyper-parameters to balance the error, the constraint $\lambda_1 > \lambda_2$ guarantees more penalty on FP ratio.

By integrating with the additional losses in adversarial feedback, the generator and critic networks are trained simultaneously as follows:

$$\mathcal{L}_\mathcal{G} = -\mathbb{E}_{\hat{c} \sim P_g}[f_c(\hat{c})] + \gamma_1 \mathcal{L}_S + \gamma_2 \mathcal{L}_W \quad (16)$$

$$\mathcal{L}_{f_c} = -\mathbb{E}_{c \sim P_r}[f_c(c)] + \mathbb{E}_{\hat{c} \sim P_g}[f_c(\hat{c})] \quad (17)$$

where $\mathcal{L}_\mathcal{G}$ and $\mathcal{L}_{f_c}$ are the objective functions for $\mathcal{G}$ and $f_c$ respectively, $\gamma_1 \geq 0$ and $\gamma_2 \geq 0$ are penalty parameters on supervised loss and weighted loss respectively.

To guarantee the Lipschitz constraint on the critic, it is necessary to force the weights of the critic to lie in a compact space. In the training process, we clip the weights to a fixed interval $[-\alpha, \alpha]$ after each gradient update.

### 4.4. Training Process

Figure 2 details the learning process for the generator $\mathcal{G}$ and critic $f_c$ in the *IndexGAN*. Denote $\mathcal{D}_{train}$ as the training data containing input features $X$ and real sequence $c$, and denote $\mathcal{D}_{batch}$ as a subset of $\mathcal{D}_{train}$ for the training in each batch. We first initialize $\mathcal{G}$ and $f_c$. In each training iteration, the prior noise $z$ sampled from Gaussian distribution and input features are fed into $\mathcal{G}$ to generate predicted sequence $\hat{c}$, then critic estimates Wasserstein distance between real and fake sequences. We compute the gradient from critic loss $\mathcal{L}_{f_c}$ and update parameters $\Theta_c$ of $f_c$ (line 8-9), also clip all weights to accelerate them till optimality. We keep training critic for $n_{critic}$ times, then we focus on generator training. The reason for this iterative process is that gradient of Wasserstein becomes reliable as the number of iterations for critic increases. Usually, $n_{critic}$ is set to 5. For generator training, predicted sequences are generated by $\mathcal{G}$ and fed into $f_c$. Then $\mathcal{L}_\mathcal{G}$ is calculated by Eq. 14-16 and parameters $\Theta_g$ of $\mathcal{G}$ is updated (lines 15-16). The number of epochs is set to 80. After the training in an adversarial manner, both $\mathcal{G}$ and $f_c$, with well-trained parameters are converged and used for deployment to real-world data.

---

**Input**: training data $\mathcal{D}_{train}$, Generator $\mathcal{G}$, Critic $f_c$
**Output**: well-trained models $\mathcal{G}$ and $f_c$ with optimal parameters
**Parameter**: model parameters: $\Theta_\mathcal{G}$, $\Theta_c$, penalty $\gamma_1$, $\gamma_2$, balancing parameters $\lambda_1$, $\lambda_2$, clipping threshold $\alpha$, the number of epochs $K$, the number of iterations of critic per generator iteration $n_{critic}$

1: Initialize $\mathcal{G}$, $f_c$, $i \leftarrow 0$.
2: **for** $epoch \in \{1, \ldots, n_{epochs}\}$ **do**
3:    **for** $\mathcal{D}_{batch} \leftarrow \mathcal{D}_{train}$ **do**
4:       $\mathbf{X}, \mathbf{c} \leftarrow \mathcal{D}_{batch}$.
5:       Sample a batch of $z$ from Gaussian.
6:       $\hat{\mathbf{c}} = \mathcal{G}(\mathbf{X}, z; \Theta_\mathcal{G})$.
7:       Compute $f_c(\hat{\mathbf{c}}; \Theta_c)$ and $f_c(\mathbf{c}; \Theta_c)$.
8:       Compute $\mathcal{L}_{f_c}$ by Eq.17.
9:       Update $\Theta_c$ in $f_c$ by minimizing $\mathcal{L}_{f_c}$.
10:      Clip weights of $\Theta_c$ into $[-\alpha, \alpha]$.
11:      $i \leftarrow i + 1$.
12:      **if** $i \% n_{critic} == 0$ **then**
13:         $\hat{\mathbf{c}} = \mathcal{G}(\mathbf{X}, z; \Theta_c)$.
14:         Compute critic value $f_c(\hat{\mathbf{c}}; \Theta_c)$.
15:         Compute $\mathcal{L}_\mathcal{G}$ by Eq.14-16.
16:         Update $\Theta_g$ in $\mathcal{G}$ by minimizing $\mathcal{L}_\mathcal{G}$.
17:      **end if**
18:    **end for**
19: **end for**

---

Figure 2 Training Process of *IndexGAN*

### 4.5. Deployment

Once the training is completed, the model is ready to be deployed. Recalling that generated sequence from our proposed architecture covers $q$ steps ahead, the predicted price can be at either location from 1 to $q$ in the generated sequence. Thus, we propose a rolling strategy for deployment. To predict the target at time $t'$, the model is repeated for $q$ rounds. For each round $j$, where $j = 1, \ldots, q$, the historical features in the time interval $[t' - w + 1 - j, t' - j]$ is learned to generate sequence at time $[t' + 1 - j, t' - j + q]$ where $\hat{c}_t^j$ is the predicted return at

$j^{th}$ position. Following this method and rolling the window forward, we obtain a sequence of prediction $\{\hat{c}_t^j\}_{j=1}^q$ with dynamic change. We take the average of the sequence to obtain the final predicted return $\hat{c}_{t'}$ whose direction is the final predicted movement $\hat{y}_{t'}$ at time $t'$. The rolling strategy considers the dynamic change of prediction since the trend of the stock market can be rapidly influenced by the various factors discussed earlier. Meanwhile, the averaged predicted return mitigates the noise caused by the uncertainty of the financial market. We expect the predicted horizon $q$ to be 5 which can generate the best result considering every week has five trading days, and our model evaluation verifies this conjecture.

## 5. MODEL EVALUATION

To evaluate the performance of the proposed model, extensive experiments are carried out with real-world stock markets and news data. We will first introduce implementation details including data description, baseline models, evaluation metrics, and parameter settings. Then we will show the empirical results of compared models, forecasting horizon analysis, and ablation study to demonstrate the effectiveness and generality of the proposed model, *IndexGAN*. The code is publicly available[1].

### 5.1. Datasets

Stock datasets contain the historical daily stock prices in Dow Jones Industrial Average (DJIA) and S&P 500 from Aug-08-2008 to Jul-01-2016, downloaded directly from Yahoo Finance. DJIA is one of the oldest equity indices, tracking merely 30 of the most highly capitalized and influential companies in major sectors of the U.S. market. S&P 500 is one of the most commonly followed indices and is composed of the 500 largest publicly traded companies. Both indices are used as reliable benchmarks of the economy. The chosen period covers a financial crisis and its following recovery period, which provides stability to our method. Volatility indices of the same period, VIX for S&P 500 and VXD for DJIA, are retrieved from Yahoo Finance. Both price and volatility features $s_t$ are converted to return calculated by $s_t/s_{t-1} - 1$. The employed technical indicators are calculated given the retrieved price data. The details are shown in Table 1.

News data is downloaded from Kaggle[2] and is publicly available. It is crawled from Reddit WorldNews Channel, one of the largest social news aggregation websites. Content such as text posts, images, links, and videos are uploaded by users to the site and voted up or down by other users. The news and posts are related to popular events in various industry sectors in the world, presenting an overall attitude with a high potential to influence the market. News data contains the top 25 headlines ranked by users' votes for a single day. The headlines are processed by removing stop words, punctuation, and digits from the raw text.

To tune the parameters, both stock and news data are split into training, validation, and testing sets approximately as the ratio of 8:1:1 in chronological order to avoid data leakage, as shown in Table 2. To capture the dependency of time series, a lag window with a size of $w$ time steps is moved along both stock and news data to construct sets of sequences.

Table 2. Statistics of stocks and news datasets.

| Dataset | Time spans |
|---|---|
| Training | 2008/09-2014/12 |
| Validation | 2014/12-2015/09 |
| Testing | 2015/09-2016/06 |

---

[1] https://github.com/JingyiGu/IndexGAN

[2] https://www.kaggle.com/aaron7sun/stocknews

## 5.2. Baselines

We compare the proposed *IndexGAN* with the following methods:

- **MA** [49] is one of the most popular indicators used by momentum traders. 5-day EMA is chosen.

- **ARIMA** [50] is a widely used statistical model. In our implementation, the parameters are determined automatically by the package 'pmdarima' in Python, regarding to the lowest AIC.

- **DA-RNN** [19] integrates the attention mechanism with LSTM to extract input features and relevant hidden states.

- **StockNet** [51] employs Variational Auto-Encoder (VAE) on tweets and historical prices to encode stock movements as probabilistic vectors.

- **A-LSTM** [20] leverages adversarial training to add simulated perturbations on stock data to improve the robustness of stock predictions.

- **Trans** [22] is the previous state-of-the-art model enhancing Transformer with trading gap splitter and gaussian prior to predicting stock movement.

- **DCGAN** [52] is a powerful GAN that uses convolutional-transpose layers in the generator and convolutional layers in the discriminator.

- **GAN-FD** [37] uses LSTM and convolutional layers to design a Vanilla GAN to predict stock pricing one step ahead.

- **S-GAN** [45] employs FinBERT on sentiment scores and uses Vanilla GAN with convolutional layers and GRU to predict pricing.

Note that MA and ARIMA use close prices only. DA-RNN, Adv-LSTM, Trans, DCGAN, and GAN-FD use market-based prices only, following their architecture designs. StockNet and S-GAN use both market-based features and news data.

Table 3. Parameters in *IndexGAN*.

| Parameters | DJI | SPX |
|---|---|---|
| Batch size | 32 | |
| Learning rate of generator | 0.0001 | |
| Learning rate of critic | 0.00005 | |
| GloVe embedding size $m$ | 50 | |
| Weight clipping threshold $\alpha$ | 0.01 | |
| Trade-off parameter $\lambda_1$ | 0.8 | |
| Trade-off parameter $\lambda_2$ | 0.2 | |
| Trade-off parameter $\gamma_1$ | 10 | |
| Trade-off parameter $\gamma_2$ | 3 | |
| Number of Epochs | 100 | 80 |
| Latent word vector size $g$ | 6 | 3 |
| Hidden size in Encoder | 100 | 100 |
| Hidden size in Decoder | 200 | 50 |

### 5.3. Parameter Settings

The proposed model *IndexGAN* is implemented in PyTorch and hyper-parameters are tuned based on the validation part. The length of historical sequence $w$ is set to 35 and future steps $q$ for prediction is set to 5. We use RMSprop as an optimizer. The details of the parameters in *IndexGAN* are shown in Table 3. For other baselines, we follow default hyper-parameter settings for implementation.

### 5.4. Evaluation Metrics

We evaluate the performance of all methods by three measures: Accuracy, F1 score, and Matthews Correlation Coefficient (MCC). Accuracy is the proportion of correctly predicted samples in all samples. F1 score incorporates both recall and precision. MCC takes all components of the confusion matrix into consideration, while the true negative is ignored in F1 score. Therefore, MCC is in a more balanced manner than F1 score. A low value of MCC close to 0 indicates that classes in prediction are highly imbalanced.

### 5.5. Experiment Results

#### 5.5.1. Comparison Results with Baselines

Table 4 presents the numerical performance of our *IndexGAN* and baselines on both data for trend prediction, from which we make the following observations:

1. The proposed *IndexGAN* achieves the best results on both indices regarding accuracy and MCC. On both data sets, *IndexGAN* achieves an accuracy of over 60% and MCC of around 0.200, which justifies the predictive power of our model. The previous state-of-the-art model Trans achieves the second-best performance but does not outperform *IndexGAN* on MCC. Hence, *IndexGAN* is proven to surpass all baselines, including Trans, on stock movement prediction tasks. In addition, the improvement in MCC by *IndexGAN* is more significant than accuracy, which illustrates that prediction from *IndexGAN* is in a more balanced manner.

2. Among the baselines, there exists an obvious gap between traditional time series models (MA, ARIMA) modeling on close price only and neural networks (DA-RNN, StockNet Adv-LSTM, Trans) modeling on all market-based features, demonstrating the impact of additional features and the superior ability of neural networks in capturing complex temporal dynamics for financial data. StockNet uses VAE, a generative model as well, which maps features from high dimensional space to low dimension space and infers the posterior distribution. Its performance is worse than *IndexGAN*. We postulate the reason is that it cannot learn posterior distribution well due to the large noise of stock prices. It reveals that GAN performs better in stock prediction tasks.

3. *IndexGAN* surpasses all GAN-based approaches. DCGAN is designed for image generation and exploits convolution layers. Since patterns in time series are not similar as in images, this architecture fails to capture temporal dependencies of stock features. Hence, GANs for image generation is not suitable for time series data. Moreover, fake and real sequences fed into the discriminator in GAN-FD are the same except for the last element, which is challenging for the discriminator to differentiate. The low values of MCC on both indices show that GAN-FD does not perform well on movement prediction. Furthermore, S-GAN outperforms other GAN-based baselines due to its sentiment analysis. However, the sentiment score from FinBERT in S-GAN falls into three categories (positive, neutral, and negative). The complex pattern of public emotions is hard to capture by such a simple sentiment score. The results of MCC illustrate that Vanilla GAN implemented in S-GAN prevents the high diversity of prediction due to the mode collapse problem. Therefore S-GAN falls behind *IndexGAN*. The proposed structure of *IndexGAN* achieves two times of MCC increase, in general, because it implements the concept of WGAN and multi-step prediction with horizon-wise information.

Table 4. Performance comparison based on the average and standard deviation of 10 runs. The top two baselines are traditional methods. The middle four models are recent state-of-the-art benchmarks. The bottom three are based on GANs.

| Model | DJIA | | | S&P500 | | |
|---|---|---|---|---|---|---|
| | Acc (%) | F1 | MCC | Acc (%) | F1 | MCC |
| EMA | 51.83 | 0.483 | 0.058 | 53.93 | 0.511 | 0.087 |
| ARIMA | 53.15 | 0.479 | 0.091 | 50.53 | 0.475 | 0.085 |
| DA-RNN | 56.23 ±1.12 | 0.642 ±0.007 | 0.104 ±0.021 | 56.71 ±1.53 | 0.513 ±0.010 | 0.092 ±0.013 |
| StockNet | 57.90 ±0.84 | 0.632 ±0.008 | 0.111 ±0.031 | 55.37 ±1.35 | 0.585 ±0.009 | 0.091 ±0.019 |
| A-LSTM | 58.12 ±1.05 | 0.674 ±0.010 | 0.145 ±0.018 | 57.13 ±1.20 | 0.596 ±0.011 | 0.143 ±0.022 |
| Trans | 58.87 ±1.01 | 0.676 ±0.009 | 0.179 ±0.027 | 58.30 ±1.34 | 0.610 ±0.011 | 0.176 ±0.031 |
| DCGAN | 56.79 ±0.09 | 0.622 ±0.003 | 0.082 ±0.014 | 55.85 ±1.01 | 0.495 ±0.008 | 0.121 ±0.010 |
| GAN-FD | 57.02 ±1.09 | 0.607 ±0.011 | 0.094 ±0.023 | 56.93 ±1.15 | 0.591 ±0.012 | 0.105 ±0.024 |
| S-GAN | 57.93 ±1.08 | 0.632 ±0.010 | 0.099 ±0.020 | 57.42 ±1.07 | 0.601 ±0.012 | 0.112 ±0.018 |
| *IndexGAN* | **60.85 ±0.95** | **0.713 ±0.006** | **0.208 ±0.024** | **60.00 ±1.37** | **0.616 ±0.014** | **0.199 ±0.027** |

### 5.5.2. Forecasting Horizon Analysis

We investigate the effects of the prediction horizon $q$ on the proposed *IndexGAN*. Figure 3 shows the change of accuracy and MCC on DJIA and SPX. As $q$ goes up from 2 to 5, the performance of the model shows a rising trend, since the critic can differentiate real and fake sequences as their length becomes longer. Both accuracy and MCC achieve the highest value when $q$ is 5. However, as the length of the predicted sequence exceeds 5 days and continues increasing to 10 days, the quality of the model degrades. This reveals that it becomes harder to capture movement patterns if the forecasting length is too long. Hence, there exists a trade-off between the uncertainty of the stock market and the ability of GAN. Choosing the appropriate prediction horizon is critical to achieving significant performance. The experiment conducted on various lengths justifies our conjecture that 5 days forward is most effective because it is a normal and popular short trading period.

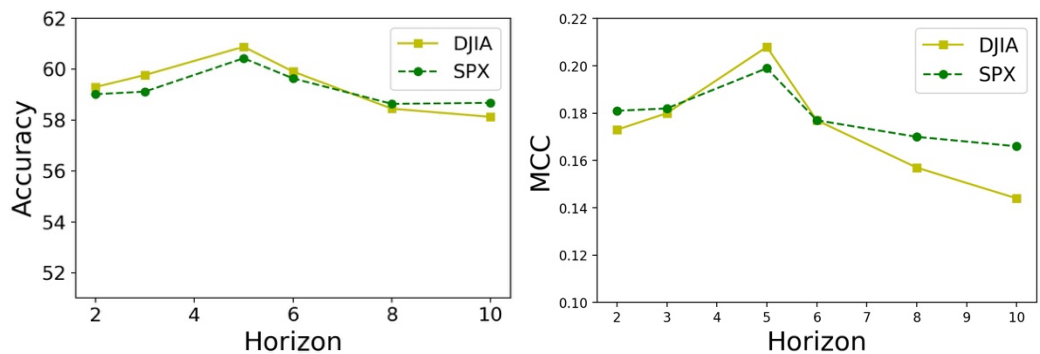

Figure 3 Influence of different settings on predicted horizon

### 5.5.3. Ablation Study

We conduct experiments on ablation variants of *IndexGAN* to fully examine the predictive power gained from each component:

- **IndexGAN-Block**: Blocks in news context learning are removed. The GloVe embedding matrix for each headline is flattened and concatenated with price features as input to be fed into seq2seq learning.

- **IndexGAN-News**: News headlines are removed from input features. The noise and market-based features are concatenated together and fed into attentive seq2seq learning directly. Thus, news context learning is removed as well.

- **IndexGAN-Vol**: The volatility index is removed from input features.

- **IndexGAN -WDist**: Wasserstein distance is removed and Vanilla GAN is implemented.

- **IndexGAN-Attn**: The attention layer in seq2seq learning is removed. The output from GRU in the encoder is directly presented as input to the decoder.

- **IndexGAN-Loss**: $\mathcal{L}_S$ and $\mathcal{L}_W$ are removed and only adversarial loss is used.

Figure 4 exhibits the performance of variants. Firstly, IndexGAN-News, IndexGAN-Block, and IndexGAN-FC are employed to test the importance of news context learning. With market-based features only, the performance of IndexGAN-News decreases significantly. It validates that news reflects the information in the stock market and can efficiently boost performance. On the other hand, even with news, the performance of IndexGAN-Block is not ideal. It illustrates our notion that directly passing the embedding matrix into the seq2seq model is likely to mask the impact of price. When the embedding size is thousands of times greater than the price, the encoder and decoder concentrate on embedding entirely. This phenomenon reveals that merely depending on the news is challenging, and both price and news features are necessary for stock movement prediction. Compared with a single fully connected layer, the blocks containing nonlinear layers bring approximately a 3% accuracy increase on the S&P 500 and DJIA, and almost 0.05 of MCC increase on the S&P 500. This phenomenon justifies that nonlinear layers in the designed blocks can boost the predictive power efficiently. With the help of blocks that fully investigate the news context and shrink the size of the embedding matrix gradually, *IndexGAN* achieves the highest results.

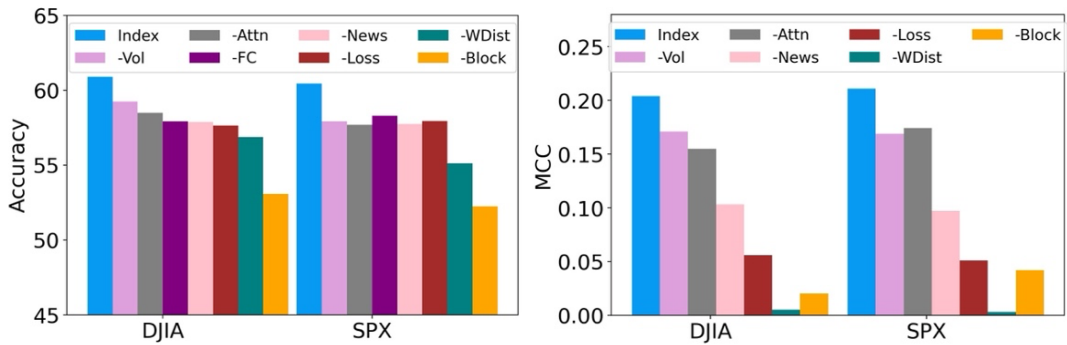

Figure 4 Ablation study results on Accuracy (top) and MCC (bottom) for comparing variants of *IndexGAN*. For example, -Attn represents *IndexGAN* without attention in seq2seq learning.

Secondly, IndexGAN-WDist does not perform well, especially MCC is close to 0 which is even worse than EMA, revealing low diversity of predicted movements and the mode collapse problem of IndexGAN-WDist. Simply leveraging Vanilla GAN cannot capture dynamics for movement prediction. It's necessary to employ the concept of Wasserstein distance to measure the similarity of prediction and ground truth, which improves the quality of the model significantly by accelerating the speed of convergence and avoiding mode collapse.

Thirdly, without the volatility index as a feature, the accuracy of Index-Vol on the S&P 500 decreases by 2.5% which is 1% more than the drop in accuracy on DJIA. The results indicate the necessity of a volatility index, especially for VIX which depends on more options than VDX.

Moreover, by comparing the IndexGAN-Attn and IndexGAN-Loss with *IndexGAN*, the attention mechanism and two additional losses on the generator enhance the predictive power by over a 2% accuracy increase.

## 6. CONCLUSION

In this paper, we present a new formulation for predicting stock movement. This work is first to propose the use of Wasserstein Generative Adversarial Networks for multi-step stock movement prediction which includes particularly crafted designs for inherent characteristics of the stock market. Specifically, in the generator, *IndexGAN* first exploits prior noise to improve the robustness of the model, implements news context learning to explore embeddings from emerging news, and employs volatility index as one of the impacting factors. Then an attentive encoder-decoder architecture is designed for seq2seq learning to capture temporal dynamics and interactions between market-based features and word representations. The critic is used for approximating Wasserstein distance between the actual and predicted sequences. Moreover, *IndexGAN* adds two additional losses for the generator to constrain similarity efficiently and avoid biased movement during optimization. Also, to mitigate uncertainty, a rolling strategy for deployment in employed. Experiments conducted on DJIA and S&P 500 show the significant improvement of *IndexGAN* and the effectiveness of multi-step prediction over other state-of-art methods. The ablation study verifies the significant advantage of each component in the proposed *IndexGAN* as well. Given this, in future research, we will extend our work to other chaotic time series where it is difficult to predict future performance from past data, especially those areas driven by human factors (i.e., commerce market).